\documentstyle[psfig,times]{mn}
\begin{document}
\hsize=6truein

\def\bfw{{\bf w}}

\def\ltsima{$\; \buildrel < \over \sim \;$}
\def\simlt{\lower.5ex\hbox{\ltsima}} \def\gtsima{$\; \buildrel > \over
\sim \;$} \def\simgt{\lower.5ex\hbox{\gtsima}}

\renewcommand{\thefootnote}{\fnsymbol{footnote}}

\title[`Hyper Parameters' Approach to Joint Estimation]
{`Hyper Parameters' Approach to Joint Estimation: Applications
to Cepheid-Calibrated Distances and X-Ray Clusters}

\author[Pirin Erdogdu, Stefano Ettori, Ofer Lahav] 
{\parbox[]{6.in} {Pirin Erdogdu$^{1,2}$, Stefano Ettori$^3$ and
Ofer Lahav$^1$\\ \footnotesize $^1$Institute of Astronomy, Madingley
Road, Cambridge CB3 0HA, UK \\ $^2$Department of Physics, Middle East
Technical University, 06531, Ankara, Turkey  \\ $^3$ESO,
Karl-Schwarzschild-Str. 2, 
D-85748 Garching bei Munchen, Germany }} \maketitle \date{February
2002}

\begin{abstract}
We use a generalised
procedure for the combined likelihood analysis of different
cosmological probes, the `Hyper-Parameters'
method, that allows freedom in the relative weights of the raw
measurements. We perform a joint analysis of the cepheid-calibrated
data from the Hubble Space Telescope Key Project and the baryon mass fraction
in clusters to constrain the total matter density of the universe,
$\Omega_{\rm m}$, and the Hubble parameter, $h$. We compare the results
obtained using Hyper-Parameters method with 
the estimates from standard $\chi^{2}$ analysis. We assume that the
universe is spatially flat, with a cosmological constant. We adopt the
Big-Bang nucleosynthesis constraint for the baryon density, assuming
the uncertainty is Gaussian distributed. Using this and the cluster
baryon fraction data, we find that the matter density and the
Hubble constant are correlated, $\Omega_{\rm m} h^{0.5} \approx 0.25$,
with preference for a very high $h$. To break the degeneracy, we add
in the cepheid-calibrated data and find the best fit values
($\Omega_{\rm m}$, $h$) =
($0.26^{+0.06}_{-0.06}$, 
$0.72^{+0.04}_{-0.02}$) (68 per cent confindence limits) using the
Hyper-Parameters approach. We use the derived Hyper-Parameters to
`grade' the 6 different data sets we analyse. Although our analysis is
free of assumptions about the power spectrum of fluctuations, our
results are in agreement with the $\Lambda$-Cold Dark
Matter `concordance' parameters derived from the Cosmic Microwave
Background anisotropies combined with Supernovae Ia, redshift surveys
and other probes.

\end{abstract}


\section{INTRODUCTION}

Combining different cosmological observations has become an essential
and common approach in cosmology. A number of groups (e.g. Gawiser \&
Silk 1998; Webster {\it et al.} 1998; Lineweaver 1998; Eisenstein, Hu
\& Tegmark 1999; Efstathiou {\it et al.} 1999, 2002; Bridle {\it et
al.} 1999, 2001a; Bahcall {\it et al.} 1999; Lahav {\it et al.} 2002)
investigated a range of cosmological parameters by joint analysis of
various cosmic probes, e.g. the Cosmic Microwave Background (CMB),
Supernovae Ia (SNIa) and redshift surveys.

It is well known that a simultaneous analysis of different probes is
essential in finding tight constraints on the cosmological parameters
and breaking the intrinsic degeneracies inherent in any single
measurement. While this is true, joint likelihood analyses pose
several statistical problems. One of these problems arise
when data 
sets are in disagreement  (e.g. Press 1996). In this case, the general
approach is to exclude the inconsistent measurements in a somewhat
ad-hoc way. A more 
objective approach to this problem was presented in Lahav {\it et al.}
(2000) and Hobson {\it et al.} (2002), who generalised the
conventional joint analysis by utilizing 
`Hyper Parameters' (hereafter HPs). The HPs provide a useful
diagnostic in determining the relative weight that should be given to
each experiment. Thus, this procedure gives an objective understanding
as to which measurements are problematic and need further assessment
of the random and systematic errors. The formalism
for the HPs is given in Appendix B.

In this paper we focus on $\Omega_{\rm m}$, the ratio of matter
density to the critical value necessary to close the Universe, and the
Hubble parameter, $H_0 = 100 h$ km s$^{-1}$ Mpc$^{-1}$.  We first
consider the measurements separately and then combine them using the
HPs method, as well as the conventional $\chi^{2}$ analysis. We assume a
flat universe, $\Omega_{\rm m} + \Omega_{\Lambda} = 1$, in agreement
with the latest CMB results (e.g. de Bernardis {\it et al.} 2002). We
adopt the Big-Bang nucleosynthesis value for $\Omega_{\rm b}h^2$,
where $\Omega_{\rm b}$ is the ratio of baryonic matter density to the
critical density.

We investigate the Hubble Space Telescope Key Project
cepheid-calibrated data (Freedman {\it et al.} 2001) and estimates of
gas mass fraction, $f_{\rm gas}$, in clusters of galaxies obtained
from X-ray observations (Ettori \& Fabian 1999; Mohr, Mathiesen \& 
Evrard 1999; Allen, Schmidt \& Fabian 2002; following the earlier work of White {\it et al.} 1993).
Unlike analyses (e.g. of the CMB and redshift surveys) that assume a
power spectrum of fluctuations in a particular scenario of dark
matter, we have selected two probes that when combined measure
$\Omega_{\rm m}$ with minimal assumptions, independent of the nature
of dark matter. The main assumptions we made in our analysis are as
follows: (i) the local $H_0$ measurements are typical of the entire
universe, (ii) the clusters of galaxies are representitive of the
matter distribution on large scales (but see e.g. Bahcall \& Comerford
2002), (iii) all the systematic errors of the measurements are
included.

In Sections 2 and 3, we introduce the data and discuss the methods
used to produce likelihoods for the individual data sets. We combine
the two data sets in Section 4.  In Section 5, we discuss our results
and compare them with other studies of baryon fraction and independent
measurements from the CMB and the 2dF galaxy redshift survey.

\section{Hubble Constant from Cepheid-Calibrated Distances}
One of the most important results on the Hubble constant comes from
the Hubble Space Telescope Key Project (Freedman {\it et al.}  2001,
hereafter F01). The group has used the cepheid period-luminosity
relations to obtain distances to 31 galaxies and calibrated a number
of secondary distance indicators measured over distances of 400 to 600
Mpc.  The values of $h$ derived using these methods and the
uncertainties (random (r) and systematic (s)) are summarised in Table
1.  Combining the measurements in Table 1 by several statistical
methods they derive the final result as $h$ = 0.72$\pm$ 0.03$_r$ $\pm$
0.07$_s$. Given the importance of this widely quoted result, we
perform a more principled statistical analysis than used by F01 to test
robustness of this value.

\begin{table}
\begin{tabular}{@{}lrl@{}}
\hline Sample & $h$ & Error \\ \hline 36 Supernovae Ia (SNIa) & 0.71 &
$\pm$ 0.02$_r$ $\pm$ 0.06$_s$ \\

21 Tully-Fisher Clusters (TF) & 0.71 & $\pm$0.03$_r$ $\pm$ 0.07$_s$ \\

11 Fundamental Plane Clusters (FP) & 0.82 & $\pm$ 0.06$_r$ $\pm$
0.09$_s$ \\

Surface Brightness Fluctuations (SBF) for 6 Clusters & 0.70 & $\pm$
0.05$_r$ $\pm$0.06$_s$ \\ \hline
\end{tabular}
\label{table:cepheids}
\caption{The values of $h$ and the 1-sigma random(r) and systematic(s)
uncertainties for Cepheid-calibrated samples (from F01, and the
references therein)}.
\end{table}

We use the raw data given in the tables in F01 for Surface Brightness
Fluctuations (SBF), SNIa, Tully-Fisher (TF) and Fundamental Plane
(FP). We combine these using the standard joint $\chi^2$ and then the
HPs approach. Each method is affected by both systematic errors which
are common to all of the methods (e.g. the adopted distance modulus to
the Large Magellanic Cloud, metallicity dependence of cepheid
period-luminosity relation and reddening by dust) and systematic
errors which are specific to each method.  In principle, some of
systematic errors can be modelled and incorporated in the $\chi^2$
analysis by adding extra parameters and marginalising over them (for
an application to CMB data see Bridle {\it et al.} 2002). However, we
follow F01 and for the variance, $\sigma^2$, in the $\chi^2$ analysis
we use the quoted random and systematic errors (from Table 1), added
in quadrature.  We also test the results when we assume random errors
only (see Table 3).  In both cases we find the best fit $h$ = 0.72
with the HPs method. This value is in very good agreement with the HST
Key project result and also with the standard $\chi^2$ analysis we
performed (Table 2).

\begin{figure}
\psfig{figure=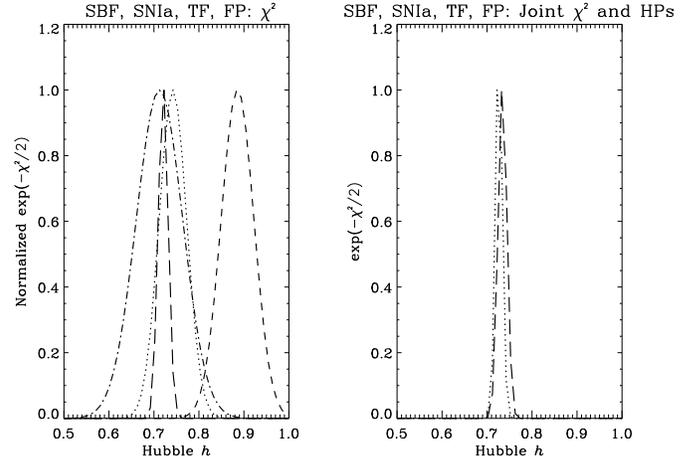,width=0.5\textwidth,angle=0}
\caption[]{ Probability functions for the Hubble constant.  The left
plot shows the $\chi^2$ statistic for four Cepheid-calibrated distance
indicators from F01: SBF(dashed-dotted line),
SNIa(long-dashed),TF(dotted) and FP(dashed).  The right plot shows the
probabilities based on the joint $\chi^2$ (long-dashed) and the HPs
approach (dotted) for the Hubble constant given the same data, with
peak values of $h=0.72$ and $h=0.73$, respectively. See also Lahav
2001b.}
\label{fig:hubble}
\end{figure}

The HP values we obtain for each case are given in Table 3.  As can be
seen from the HPs formalism in Appendix B, the HPs can be interpreted
as either an indication of misfit of the data set and the model
(e.g. due to systematic problems with the data or to the wrong model),
or as a rescaling of the quoted error bars (to $\sigma/\alpha^{1/2}$).
The HPs scores rank the quality of the methods as follows: SBF (the
highest), TF, SNIa and FP.  This is in accord with comments on the
precision and systematics of each method given by F01. The FP
measurements alone give the most discrepant value of $h$
(Figure~\ref{fig:hubble} and Table 2),
and they got the lowest HP (even when the systematic errors were not
included in the analysis; see Table 3). This is an example where HPs
flag a problem with the measurements. Although the discrepancy of the
$h$ value obtained 
using FP data alone does not affect the final value derived by joint
analyses of FP, SNIa, TF and SBF, one needs to be careful about using
the standard $\chi^2$. A $\chi^2$ analysis of discrepant results may
lead to a wrong joint answer. A clear example of this problem arises
if we use only the FP and the TF data, which have the most
similar scatter  (see Figure~\ref{fig:hubble}) and $\chi^2$
values (see Table 2) out of all the probes. We find that the joint
likelihood function for $h$ peaks at 0.76 using HPs analysis, whereas the best
value of $h$ is 0.80 using the  $\chi^2$ analysis. In other words, HPs
analysis 
grades the TF measurements as more reliable than the FP data and hence
gives the joint value of $h$ closer to the best value obtained using TF
measurements. Using standard joint analysis, the joint likelihood function
peaks between the likelihood functions of the two probes, as expected.
It is also interesting to note
that indeed F01 attached to the FP data the largest systematic errors
(see Table 1).  

As can be seen in Figure~\ref{fig:hubble}, the joint probability
functions obtained using the two approaches ($\chi^2$ and HPs) are
both dominated by SNIa data, since SNIa as an indicator possesses the
smallest random scatter.

\begin{table}
\begin{tabular}{@{}lrrr}
\hline Sample & $\chi^2$ & Best $h$ & 68 per cent confidence limits \\
\hline 36 SNIa & 19.0 & 0.72 & $0.70 \,<\, \ h \,<\, 0.73$ \\

21 TF & 7.6 & 0.74 & $0.70 \,<\, \ h \,<\, 0.76$ \\

11 FP & 4.9 & 0.88 & $0.84 \,<\, \ h \,<\, 0.92$ \\

6 SBF & 1.7 & 0.72 & $0.66 \,<\, \ h \,<\, 0.76$ \\ \hline Joint $\chi^2$ (for 74 data points) &
57.0 & 0.73 & $0.70 \,<\, \ h \,<\, 0.77$ \\ \hline
\end{tabular}
\label{table:derived}
\caption{ The conventional $\chi^2$ analysis using the raw F01
data. For each sample the best fit value of $h$ and the $\chi^2$ value
at this point are given.}
\end{table}

\begin{table}
\begin{tabular}{@{}lcc}
\hline Sample & HPs & HPs \\ & (r only) & (r+s) \\ \hline 36 SNIa &
0.3 & 1.9\\

21 TF & 1.9 & 2.7 \\

11 FP & 0.2 & 0.5 \\

6 SBF & 2.8 & 3.4 \\ \hline Best $h$ & 0.72 & 0.72 \\ \hline  68 per
cent confidence limits  & $0.69 \,<\, \ h \,<\, 0.76$ & $0.68 \,<\, \ h \,<\, 0.77$ \\ \hline 
\end{tabular}
\label{table:derived}
\caption{The Hyper-parameters analysis using the raw F01 data. For
each case the HP is given. The first column is shows the HPs obtained
using random errors only, while the second column is for
random+systematic errors, added in quadrature.}
\end{table}

\section{X-Ray Observations of Clusters of Galaxies}

Clusters of galaxies provide various means for the study of the
cosmological parameters. Clusters are the largest gravitationally
bound structures in the Universe, and therefore generally are assumed to
be the tracers of matter distribution on large scales. Provided that
$\Omega_{\rm b}h^2$, the baryon density, can be inferred from
primordial nucleosynthesis abundance of the light elements, the
cluster baryon fraction, $f_{\rm b}$ = $\frac {\Omega_{\rm b}}
{\Omega_{\rm m}}$, can then be used to constrain $\Omega_{\rm m}$ and
$h$ (White {\it et al.} 1993, Steigman, Hata \& Felten 1999, Ettori
2001). The baryons in clusters are primarily in the form of X-ray
emitting gas that falls into the cluster halo and secondarily in the
form of stellar baryonic mass. Hence the baryon fraction in clusters is
estimated to be

\begin{equation}
f_{\rm b} = \frac{\Omega_{\rm b}} {\Omega_{\rm m}} \simgt f_{\rm gas}
+ f_{\rm gal},
\label{eq:fbar}
\end{equation}
where $f_{\rm b} = M_{\rm b}/M_{\rm grav}$, $f_{\rm gas} = M_{\rm
gas}/M_{\rm grav}$, $f_{\rm gal} = M_{\rm gal}/M_{\rm grav}$ and
$M_{\rm grav}$ is the total gravitating mass.

We consider two different datasets of gas mass fraction estimates. The
first was published in Ettori and Fabian (1999) (hereafter EF99). This
is a sample of 36 relaxed galaxy clusters\footnote{In our analysis, we
use 35 clusters instead of 36. 
Cluster A3888 is excluded from the sample since this cluster's gas
temperature was not obtained from X-ray observations but from  the gas
temperature-optical velocity dispersion relation.} with high X-ray luminosities
($L_{\rm X} \ga 10^{45}$ erg s$^{-1}$) and a redshift range from 0.05
to 0.44. The second data
set is taken from Mohr, Mathiesen and Evrard 
(1999) (hereafter MME99). This is an X-ray flux-limited sample of 45
clusters with redshifts between 0.01 and 0.18.  Both groups use ROSAT
PSPC surface brightness profiles and intercluster medium (ICM)
temperatures from ASCA, Ginga and Einstein MPC observations for their
analyses.  The gas mass fraction estimates, $f_{\rm gas}$, in both
datasets are derived assuming the gas is isothermal and in hydrostatic
equilibrium within the limiting radius, $R_{\Delta}$, where $\Delta$
is the mean overdensity of the total mass in a cluster relative to the
background value.

To allow for the variation from cluster to cluster, for the present
analysis we use the individual $f_{\rm gas}$ per cluster, rather than
a global average.  As the clusters are at relatively high redshifts,
$f_{\rm gas}$ is a function of $\Omega_{\rm m}$ and $h$ (see Figure
2).This is due to two
cosmological effects (see discussion in, e.g. Ettori 2001).

The first dependence is due to the angular diameter distance, $d_{\rm
ang}$, which is a function of $h$ and $\Omega_{\rm m}$ in a flat
universe with cosmological constant. For an isothermal gas in
hydrostatic equilibrium, $f_{\rm gas}$ can be calculated through the surface brightness profile and gas
temperature (cf. Ettori \& Fabian 1999; Mohr, Mathiesen \& Evrard
1999). If the surface brightness profile is measured as the
integration of the thermal bremsstrahlung emissivity of the ICM along
the line of sight, one finds that $f_{\rm gas}$ is proportional to
$d_{\rm ang}^{1.5}$ (see Appendix A). We calculate the angular
diameter distances using Eq.~\ref{eq:dang_k0} and vary $h$ and
$\Omega_{\rm m}$. We find that $f_{\rm gas}$
decreases significantly (about 40 per cent) in a low density universe
with $h>0.5$.

\begin{figure}
\psfig{figure=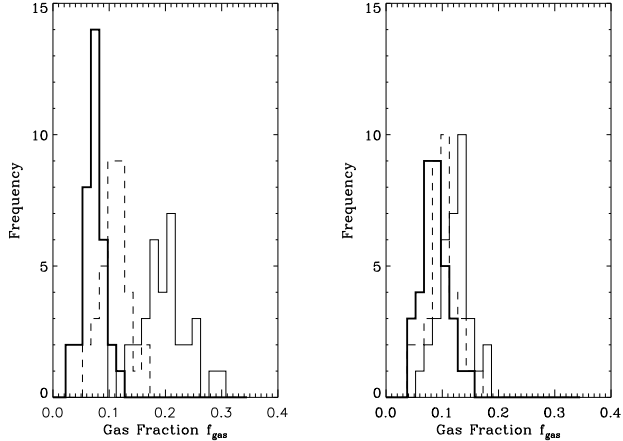,width=0.5\textwidth,angle=0}
\caption[]{ The histograms of the gas mass fraction $f_{\rm b}$
observed in 35 clusters (EF99). The left panel shows the dependence on
$h$ for fixed $\Omega_{\rm m} = 0.3$: $h = 0.5$ (solid line), $h =
0.75$ (dashed line) and $h = 1.0$ (thick solid line). The right panel
shows the dependence on $\Omega_{\rm m}$ for fixed $h = 0.75$:
$\Omega_{\rm m} = 0.1$ (solid line), $\Omega_{\rm m} = 0.5$ (dashed
line) and $\Omega_{\rm m} = 1.0 $ (thick solid line).}
\label{fig:histogram}
\end{figure}

The second cosmological dependence is weaker. The hydrodynamical
simulations (see e.g. Evrard {\it et al.} 1996) show that for an
Einstein-de Sitter cosmology, a mean overdensity, $\Delta = 500$
defines a region within $R_{\Delta}$ where the assumptions of
isothermal gas in hydrostatic equilibrium are valid. Therefore, in
both data sets, $f_{\rm gas}$ has been estimated at $\Delta = 500$.
However, $\Delta$ is function of the cosmological parameters and
increases as $\Omega_{\rm m}$ decreases in a universe with positive
$\Omega_{\Lambda}$. $R_{\Delta}$ is proportional to $(\Omega_{\rm
m}\Delta)^{-0.5}$ (Eq.~\ref{eq:rpropDeltaO_m} in Appendix A) and
combining this with the radial dependence of the gas mass fraction
near $R_{\Delta}$ ($f_{\rm gas} \propto r^{0.2}$), we get $f_{\rm gas}
\propto(\Omega_{\rm m}\Delta)^{-0.1}$.  Figure~\ref{fig:histogram}
illustrates the dependence of $f_{\rm gas}$ on $h$ and $\Omega_{\rm
m}$. We see that $f_{\rm gas}$ is mainly sensitive to $h$.

We also include the stellar contribution in cluster galaxies, $f_{\rm
gal}$, to the baryon fraction estimate.  We use a global correction
$f_{\rm gal}$ = $(0.01\pm0.005)h^{-1}$ (Fukugita {\it et al.} 1998),
which is based on mass-to-light ratio, $M/L = 4.5 \pm1$ per galaxy,
after averaging different galaxy types, with no dependence on $h$,
hence the $h$-dependence in $f_{\rm gal}$ comes solely from the $h$
dependence of $M_{\rm grav}$ on the cluster distances.  We note that
some estimates of $M/L$ per galaxy depend on the distances to
galaxies, and hence on $h$, resulting in $h$-independent $f_{\rm gal}$
(e.g. Wang {\it et al.} 1999 quote $f_{\rm gal} = 0.013 $).  Ideally,
$f_{\rm gal}$ should be estimated per cluster, but these data are
currently unavailable, and in any case it is only a small contribution
to the total baryon fraction.

We vary the matter density of the universe in units of the critical
density, $\Omega_{\rm m}$, and the present day value of the reduced
Hubble constant, $h$, over the parameter spaces [0.01, 0.5] and [0.5,
2.5], respectively. We also put a prior on $\Omega_{\rm
b}h^2$ so that $\Omega_{\rm b}h^{2}$ = 0.019$\pm$0.002, 95\%
confidence levels, (e.g. Burles $\it et$ $\it al.$ 2001) and
marginalise over the uncertainty range in order to get the likelihoods
of $\Omega_{\rm m}$ and $h$. We note that fixing $\Omega_{\rm b}h^2$
to 0.019 tightens the constrains in the $\Omega_{\rm m}$, $h$ plane
but does not change the best fit points significantly.

We compute the $\chi^2$s using Eq.~\ref{chi2_sum} in Appendix B:

\begin{equation}
\chi^2 = \sum_{i} {\frac {1} {\sigma_{i}^{2}}} [f_{{\rm b},i}(\Omega_{\rm
m},h) - {\frac{\omega_{\rm b}}{\Omega_{\rm m}h^{2}}}]^2,
\label{chi2_clusters}
\end{equation}
where the sum is over the number of clusters, the baryon fraction is
given by $f_{\rm b} = f_{\rm gas} + f_{\rm gal}$ and $\omega_{\rm b}$
= $\Omega_{\rm b}h^{2}$.

To obtain a qualititive understanding we consider Eq.~\ref{eq:fbar}
and use the first cosmological dependence that dominates over the
others to an approximate relation between $\Omega_{\rm m}$ and $h$:
\begin{equation}
\Omega_{\rm m} = \frac{\omega_{\rm b} \ h^{-2}} {f_{\rm gas} + f_{\rm
gal}} \approx \frac{\omega_{\rm b} \ h^{-2}} {0.08 h^{-1.5} + 0.01
h^{-1}}.
\label{envelope}
\end{equation}
For the observed distribution of the gas mass fraction of about 0.08
$h^{-1.5}$ (e.g. EF99) and reasonable values of Hubble constant, the
above relation can be estimated as $\Omega_{\rm m} h^{0.5} \approx
0.25$.

\begin{figure}
\psfig{figure=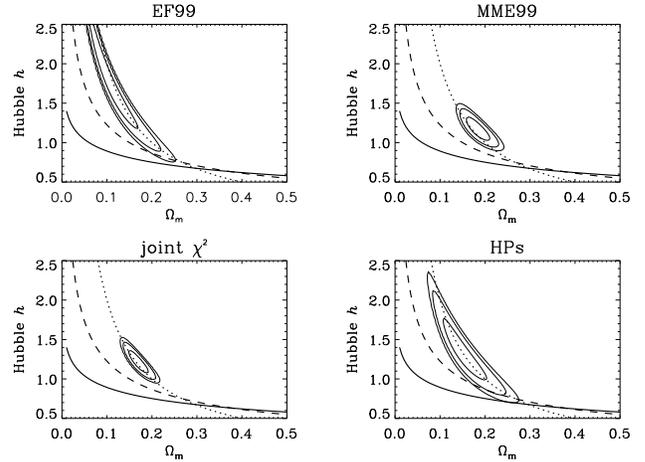,width=0.5\textwidth,angle=0}
\caption[]{ The X-ray Cluster Likelihood Functions in the
\{$\Omega_{\rm m}$,$h$\}-plane, after marginalisation over
$\Omega_{\rm b}h^2$.  The results are compared to estimates from the
2dF Galaxy Redshfit Survey: $\Omega_{\rm m}h$ $\approx$ 0.2 (dashed
lines; Percival {\it et al.} 2001) and from the CMB: $\Omega_{\rm
m}h^2$ $\approx$ 0.15 (dotted lines; Netterfield {\it et al.} 2002 and
Pryke {\it et al.} 2002) and the age of the universe $t_0$ $\approx$
14 Gyr (solid lines; Knox {\it et al.}  2001). The contours denote the
68, 95 and 99 percent confidence regions.}
\label{fig:clusters}
\end{figure}

The two datasets do have some clusters in common, however, we see that
our conclusions are not affected if we consider the samples
separately. The results are plotted in Figure~\ref{fig:clusters}. It
is seen that in both data sets the value of $h$ is unacceptably high
at the $68\%$ confidence level (but still lower than Hubble's original
value of 500 km/sec/Mpc!) and the best fit values of $\Omega_{\rm m}$
are relatively low.  The parameter values ($\Omega_{\rm m}$, $h$) at
the best fit points (with 68 per cent confidence limits) are
($0.11^{+0.03}_{-0.04}$,$1.73^{+0.33}_{-0.48}$) and
($0.18^{+0.02}_{-0.02}$,$1.17^{+0.10}_{-0.10}$) for EF99 and MME99
data, respectively. The standard joint analysis of the data sets
yields ($\Omega_{\rm m}$, $h$) =
($0.17^{+0.01}_{-0.02}$,$1.23^{+0.08}_{-0.12}$) and with HPs method we 
obtain ($\Omega_{\rm m}$, $h$) = ($0.15^{+0.03}_{-0.03}$,
$1.33^{+0.22}_{-0.28}$). It is easy to see from 
Figure~\ref{fig:clusters} that both separate and joint analyses of the
data imply that $\Omega_{\rm m}$ = 1 is ruled out at very high
confidence level. The standard $\chi^2$ and HP analyses, shown on
bottom left and right panels in Figure~\ref{fig:clusters} yield
slightly different results.  The HP values are 0.6 for the EF99 and
0.1 for the MME99 sample.  The low weight given to the MME99 data may
indicate possible systematic errors, an underestimation of the random
errors or incomplete modelling.  The HPs obtained suggest that the
EF99 sample is more reliable than the MME99 data. The galaxy clusters
in the EF99 sample were selected for their high X-ray luminosity and
relaxed morphology, whereas the selection criteria in MME99 was to
build a flux-limited sample.  Therefore, we see at least two effects
that can make the EF99 sample more robust in providing a stable
central value of the gas fraction: (i) the systematics due to
non-homogeneous objects are more under control and (ii) the observed
dependence of the value of gas mass fraction upon the plasma
temperature (and luminosity, as consequence of the tight $L-T$
relation observed in galaxy clusters; e.g. Ettori, Allen \& Fabian
2001 for an application of HPs to this relation) makes the selection
in luminosity (instead of flux) a more robust way to select objects in
the upper end of the gas mass fraction distribution.

Being concerned about the high Hubble constant, we applied the
Bootstrap method (Efron 1982) to baryon mass fraction data to ensure
that there are no outlying clusters which could alter the significance
of the obtained best fit values. We created 2000 synthetic catalogs
selected from the two samples. The histograms of the best fit points
for $h$ and $\Omega_{\rm m}$ for these bootstrap realizations are in
very good agreement with the presented results.

\begin{figure}
\psfig{figure=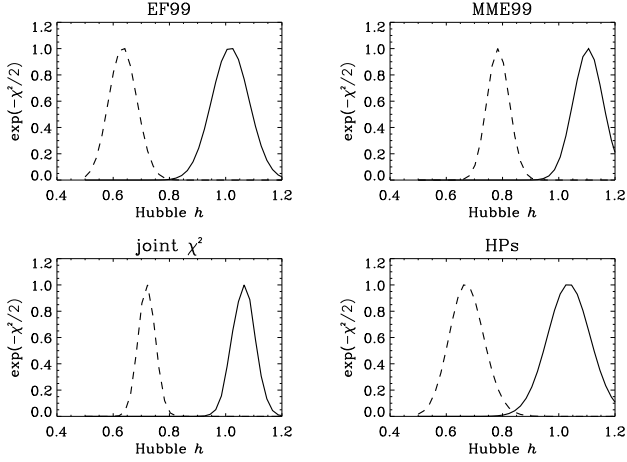,width=0.5\textwidth,angle=0}
\caption[]{ The 1-d likelihood distributions of the Hubble Constant
for fixed $\Omega_{\rm m} = 0.3 $ (dashed lines) and $\Omega_{\rm m} =
0.2 $ (solid lines) and marginalised over $\Omega_{\rm b}h^2$.  The
figure illustrates that $h$ is highly sensitive to the assumed
$\Omega_{\rm m}$, as $h$ and $\Omega_{\rm m}$ are highly correlated.}
\label{fig:clus1d}
\end{figure}

The joint analysis of cosmological probes (e.g. Bahcall {\it et al.}
1999 and Efstathiou {\it et al.} 1999) suggests a flat Universe with
$\Omega_{\rm m} \approx 0.3 $. We fix $\Omega_{\rm m}$ to 0.3 and to
0.2 and plot 1 dimensional likelihood distributions of $h$. These
plots are shown in Figure ~\ref{fig:clus1d}. For fixed $\Omega_{\rm m}
= 0.3 $ the best fit points of $h$ are 0.64 and 0.78 for EF99 and
MME99, respectively. The joint 1-dimensional likelihood distribution
peaks at $h$ = 0.72 using the standard and at $h$ = 0.66 using the HPs
approach. Note the good agreement of the best fit value of $h$ for
both data with the F01 result. When $\Omega_{\rm m}$ is fixed to be
0.2, the plots shift significantly to the right, with best fit points
$h= 1.03, 1.11, 1.07$ and 1.03 for EF99, MME99, standard joint
analysis and HPs approach, respectively.

We conclude that the baryon fraction data on its own cannot constrain
each of the two parameters, but only their combination, $\Omega_{\rm
m} h^{0.5} \approx 0.25$.  Therefore, although the likelihood peak is
at a high value of $h$ and a low value of $\Omega_{\rm m}$, we should
not attach much significance to the individual values.  To constrain
the individual values of $h$ and $\Omega_{\rm m}$ we now combine the
baryon fraction data with the cepheid sample.

\section{Combining the baryon fraction and the cepheid data}

\begin{figure}
\psfig{figure=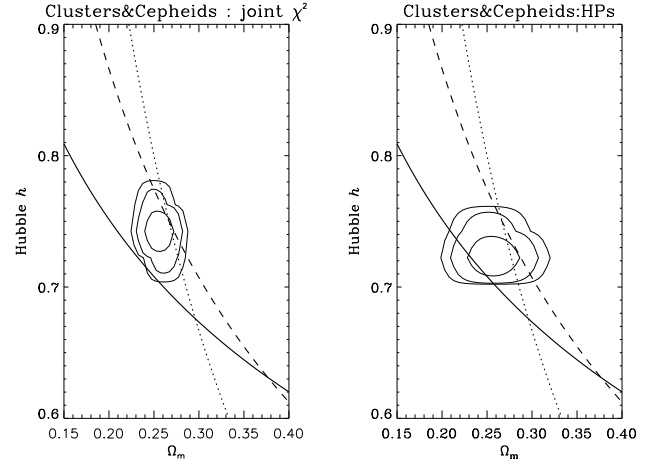,width=0.5\textwidth,angle=0}
\caption[]{ The likelihood functions from combining the Cluster data
with Cepheids.The results are compared to estimates from the 2dF
Galaxy Redshfit Survey: $\Omega_{\rm m}h$ $\approx$ 0.2 (dashed lines)
and from the CMB: $\Omega_{\rm m}h^2$ $\approx$ 0.15 (dotted lines)
and the age of the universe $t_0$ $\approx$ 14 Gyr (solid lines), as
in Figure 3. The contours denote the 68, 95 and 99 percent confidence
regions.}
\label{fig:clus_joint}
\end{figure}

We present our results for the combined analysis in
Figure~\ref{fig:clus_joint}. It can be seen that the high contour
regions obtained for $h$ with the cluster data have decreased
significantly. The confidence regions are a lot tighter than the
confidence regions of the single data sets alone, thus giving stronger
constraints.

Table 3 and Table 4 show the best fit values and $68\%$ confidence
limits of $h$ and $\Omega_{\rm m}$, respectively. The best fit points
of the parameters lie within the confidence limits, indicating that
the likelihood distributions are well behaved. The dominating data
sets are the cepheid-calibrated distance indicators. This is expected,
since the uncertainties on $h$ are much larger for the cluster
samples.

The resulting HPs of our analysis are 0.6 (EF99), 0.1 (MME99), 1.8
(SNIa), 2.7 (TF), 0.5 (FP) and 3.3 (SBF). The HPs are actually almost
identical to the HPs derived for the individual data sets in section 2
and 3.  We also see that the dominance of the cepheid data relative to
the baryon fraction data.  Since the cepheid data only constrain $h$,
this leads to a narrower error bar for $h$ in the HPs analysis
compared with the joint $\chi^2$ (which give equal weight to each of
the 6 data sets).  Since the cepheid data have no information about
$\Omega_{\rm m}$ but have higher HP values, the error bar on
$\Omega_{\rm m}$ is wider in the HPs analysis.

Another interesting result is that the HPs analysis probes a parameter
space in good agreement with the results from 2dF Galaxy Survey and
CMB anisotropy measurements. The right panel of
Figure~\ref{fig:clus_joint} shows that our combined analysis is
consistent with other measurements.

\begin{table}
\centerline{\vbox{
\begin{tabular}{@{}lcc}
\hline Analysis & Best fit value & 68 per cent confidence limits\\
\hline {\rm Joint} $\chi^2$ & $0.74$ & $0.70 \,<\, \ h \,<\, 0.78$\\
{\rm HP}s & $0.72$ & $0.70 \,<\, \ h \,<\, 0.76$\\ \hline
\end{tabular}
}}
\label{bestfit}
\caption{ The values derived for $h$ using different analysis
techniques.  The $68\%$ confidence limits are given, calculated for
each analysis by marginalising the likelihood function over
$\Omega_{\rm b}h^2$ and $\Omega_{\rm m}$.}
\end{table}

\begin{table}
\centerline{\vbox{
\begin{tabular}{@{}lcc}
\hline Analysis & Best fit value & 68 per cent confidence limits\\
\hline {\rm Joint} $\chi^2$ & $0.25$ & $0.22 \,<\, \Omega_{\rm m}
\,<\, 0.29$\\ {\rm HPs} & $0.26$ & $0.20 \,<\, \Omega_{\rm m} \,<\,
0.32$\\ \hline
\end{tabular}
}}
\label{bestfit}
\caption{ The values derived for $\Omega_{\rm m}$ using different
analysis techniques.  The $68\%$ confidence limits are given,
calculated for each analysis by marginalising the likelihood function
over $\Omega_{\rm b}h^2$ and $\Omega_{\rm m}$.}
\end{table}

\section{Discussion}

We have presented applications of a generalised procedure,
`Hyper-Parameters' (HPs), for analysing a set of different measurements. 
We performed a combined analysis of baryon fraction in clusters
and cepheid-calibrated distances.  We used the HPs
formalism for joint analyses of the data sets to constrain the
cosmological parameters $h$ and $\Omega_{\rm m}$ (assuming a flat
universe) and to check the reliability of the data sets.

Using the baryon mass fraction in clusters, we obtained a very high
best fit for $h$ . However, Figure 3 shows a strong correlation
between $h$ with $\Omega_{\rm m}$, $\Omega_{\rm m} h^{0.5} \approx
0.25$. The addition of the cepheid sample to the cluster data changed
significantly the best fit values, $h=0.72$ and $\Omega_{\rm} = 0.26$.
This is not surprising, as the accurate cepheid-calibrated distances
dominate the joint likelihood.

Recently, Douspis {\it et. al} (2001)
combined CMB data with a fixed value for baryon fraction $f_{\rm b} =
0.048 h^{-1.5} + 0.014$ and found $\Omega_{\rm m} \approx 0.4$, $h
\approx 0.6$. Their results are in agreement with our derived
combination, $\Omega_{\rm m} h^{0.5} \approx 0.25$ and in marginal
agreement with our best fit values for $\Omega_{\rm m}$ and $h$,
derived from the joint analysis of baryon fraction and the cepheid
data.  Our analysis differs from theirs in the sense that we have used
individual gas fraction values for each of the $80$ clusters in our
sample, allowing self-consistency for the dependence of the gas
fraction on $\Omega_{\rm m}$. 

More recently, Allen {\it et. al} (2002)
used gas mass fraction measurements of 6 relaxed clusters observed
with the Chandra Observatory, the HST Key Project value for $H_0$ and the
and Big-Bang nucleosynthesis value for
$\Omega_{\rm b}h^2$ to constrain $\Omega_{\rm m}$ and 
$\Omega_{\Lambda}$.  The value they obtain for $\Omega_{\rm m}$ is
again in very good agreement with our results. 

In our analysis, we have found that HPs assigned to the cluster
samples are lower than the values assigned to cepheid-calibrated
distances.  
Indeed, it is worth noticing that some systematic uncertainties affect $f_{\rm
gas}$ when it is estimated under the assumption of an isothermal gas
in hydrostatic equilibrium with the dark matter gravitational
potential, like in this paper.  Some aspects of the physics of the ICM
still need further understanding (e.g. the measurement of the total
mass changes in presence of a gradient in the temperature profile
and/or supporting pressure from non thermal component) and will be
addressed in the near future with the current X-ray missions Chandra
(Weisskopf {\it et al.} 2000) and XMM-Newton (Hasinger {\it et al.}
2001).  Moreover, we have not taken into account all contribution from
baryons in dark matter to the cluster baryon budget.

There are also significant systematic uncertainties in the HST Key Project
Result which should be addressed in the future. These
uncertainties, discussed in detail in F01, are mainly due to the errors
in the adopted distance modulus to the Large Magellanic Cloud (LMC)
upon which the Cepheid period-luminosity relation is strongly
dependent. Furthermore, the effects of reddening and metallicity on the
Cepheid period-luminosity relation maybe stronger than expected (Shanks
{\it et al.} 2002). It is also worth noting that the local variations
in the expansion rate due to large scale velocities may effect the
accurate determination of $H_0$ (eg. Turner, Cen $\&$ Ostriker 1992).

Although our analysis is free of assumptions about the power spectrum
of fluctuations, the results we obtain are in remarkable agreement
with the $\Lambda$-Cold Dark Matter `concordance' parameters (e.g
Figure 5) derived from the Cosmic Microwave Background anisotropies
combined with Supernovae Ia, 2dF galaxy redshift survey and other
probes.

\section*{ACKNOWLEDGEMENTS}

We thank Steve Allen, Sarah Bridle, Wendy Freedman, Jeremy Mould and
Jerry Ostriker for their 
helpful comments. PE acknowledges support from the Middle East
Technical University, Ankara, Turkey, the Turkish Higher Education
Council and Overseas Research Trust.

\appendix
\section{Estimation of the gas fraction}

For an isothermal plasma in hydrostatic equilibrium, the gas mass
fraction, $f_{\rm gas}$, is calculated as the ratio of the mass of the
gas and of the total gravitating mass, $M_{\rm grav}$, within the
radius, $R_{\Delta}$, at which the given mean overdensity of the total
mass within cluster, $\Delta$, with respect to the background value,
$\Omega_{\rm m}\rho_{\rm c}$, is reached:

\begin{equation}
f_{\rm gas} = \frac{M_{\rm gas} (<R_{\Delta})}{M_{\rm grav}
(<R_{\Delta})} = \frac{4 \pi \int_0^{R_{\Delta}} \rho_{\rm gas}(r) r^2
dr} { \frac{k T_{\rm gas} R_{\Delta}}{G \mu m_{\rm p}} \left(-
\frac{\partial \ln \rho_{\rm gas}}{\partial \ln r}
\right)_{r=R_{\Delta}} },
\label{eqn:fgas}
\end{equation}
where $R_{\Delta} = \theta \ d_{\rm ang}$ is the physical radius,
$\theta$ is the angular separation, $d_{\rm ang}$ is the angular
diameter distance, $T_{\rm gas}$ is the ICM temperature, $k$ is the
Boltzmann constant, $\mu$ is the mean molecular weight in
a.m.u. ($\sim 0.6$), $G$ is the gravitational constant and $m_{\rm p}$
is the proton mass.

The surface brightness, $S(\theta)$ is given by the integral along the
line of sight:
\begin{equation}
S(\theta) \propto \int\rho_{\rm gas}^2 T_{\rm gas}^{0.5} dl.
\label{eqn:surfacebrightness}
\end{equation}
Hence, the gas density, $\rho_{\rm gas}$ is proportional to $d_{\rm
ang}^{-0.5}$. Combining this with the other dependence in Eq.~\ref{eqn:fgas},

\begin{equation}
f_{\rm gas} \propto \frac{d_{\rm ang}^{3-0.5}}{d_{\rm ang}} = d_{\rm
ang}^{1.5} \propto h^{-1.5}.
\label{eqn:fgasdang}
\end{equation}

For $\Omega_{\rm k}=0$, $d_{\rm ang}$ can be written as
\begin{equation}
d_{\rm ang} = \frac{c}{H_0 (1+z)} \cdot \left\{ \begin{array}{ll} z &
 \mbox{if $\Omega_{\rm m} = 0$} \\ \int^z_0 \frac{\Omega_{\rm
 m}^{-1/2} \ d \zeta }{\left[ (1+\zeta)^3 +\Omega_{\rm m}^{-1} -1
 \right]^{1/2}} & \mbox{otherwise}
\end{array} \right.
\label{eq:dang_k0}
\end{equation}

The mean overdensity, $\Delta$, for $\Omega_{\rm k}$= 0 is given by
(e.g. Kitayama \& Suto 1996; Henry 2000)
\begin{equation}
\Delta(\Omega_{\rm m},z) = \Delta(\Omega_{\rm m}=1) \cdot \left[
1+0.4093 \left(\Omega_{{\rm m},z}^{-1} - 1 \right)^{0.9052} \right],
\end{equation}
where $(\Omega_{{\rm m},z}^{-1} -1) =(\Omega_{\rm m}^{-1}-1)/(1+z)^3$.

The observed mass profile is calculated as
\begin{equation}
M_{\rm grav}(<R_{\Delta})= (4/3) \pi R_{\Delta}^3 \Omega_{\rm m}
\rho_{\rm c} (1+z)^3 \Delta
\end{equation}
and is proportional to $R_{\Delta}$ for a virialised system so that
\begin{equation}
R_{\Delta} \propto (\Omega_{\rm m} \Delta)^{-0.5}.
\label{eq:rpropDeltaO_m}
\end{equation}

\section{Hyper-Parameters}

Lahav {\it et al.} (2000) (see also Bridle 2000, Lahav 2001a \&
2001b), generalised the standard procedure of combining likelihoods
using the 'Hyper-Parameters' approach. The conventional way of
combining likelihood functions of different data sets is either to
give all the sets the same statistical weight or to assign weights in
an ad-hoc way. The `Hyper-Parameters' method generalises this approach
by assigning each data a relative weight.

Given two independent data sets $D_{A}$ and $D_{B}$ (with $N_{A}$ and
$N_{B}$ data points respectively), our approach is to combine the
$\chi^{2}$s in the following manner:

\begin{equation}
\chi^2_{\rm joint} = \alpha \chi^2_A \; + \beta \; \chi^2_B \; ,
\label{chi_2hp}
\end{equation}
where $\alpha$ and $\beta$ are `Hyper-Parameters'(HPs).  The maximum
likelihood of a given model is estimated by minimising the above
quantity. We calculate the $\chi^2$s using the equation below:
\begin{equation}
\chi^2 = \sum {\frac {1} {\sigma_{i}^{2}}} [x_{{\rm obs},i} - x_{{\rm
pred},i}(\bfw)]^2\;
\label{chi2_sum}
\end{equation}
where the sum is over the number of measurements, $\sigma_{i}$ is the
error for each data point and $\bfw$ is the vector of free parameters
we wish to determine (e.g. $\Omega_{\rm m}$ and $h$).

The HPs are eliminated by marginalisation over $\alpha$ and $\beta$:
\begin{equation}
P(\bfw| D_{A}, D_{B}) = \int \int P(\bfw, \alpha, \beta | D_{A},
D_{B}) \;d \alpha \;d \beta \;.
\end{equation}

In order to evaluate the above integral, we use Bayes' theorem to
write the following relations:
\begin{equation}
P(\bfw, \alpha, \beta | D_{A}, D_{B}) = \frac {P(D_{A}, D_{B} | \bfw,
\alpha, \beta) \;P(\bfw, \alpha, \beta) } {P(D_{A}, D_{B})} \;,
\label{eqn:bayes}
\end{equation}
and
\begin{equation}
P(\bfw, \alpha, \beta) = P(\bfw | \alpha, \beta) \; P(\alpha, \beta)
\;.
\end{equation}
We also assume the following:
\begin{equation}
P( D_{A},D_{B}|\bfw, \alpha, \beta) = P(D_{A}| \bfw,\alpha) \;
P(D_{B}|\bfw,\beta) \;,
\end{equation}
\begin{equation}
P(\bfw |\alpha, \beta) = const. \;,
\end{equation}
and
\begin{equation}
P(\alpha, \beta) = P(\alpha) \; P(\beta) \;.
\label{eqn:prior}
\end{equation}
We take the prior probabilities in Eq.~\ref{eqn:prior} as Jeffreys`
uniform priors in the log, $P(\ln \alpha) = P(\ln \beta) =1$. Assuming
Gaussianity, we write $P(D_{A} | \bfw, \alpha) \; \propto \;
\alpha^{N_{A}/2}\; \exp (-{\alpha \over 2} \chi_{A}^{2} )$ and
similarly for $D_{B}$.  It then follows that the probability for the
parameters $\bfw$ given the data sets is:
\begin{equation}
-2 \; \ln P(\bfw| D_{A}, D_{B})\; = N_{A} \ln (\chi_{A}^{2}) \; + \;
N_{B} \ln (\chi_{B}^{2})\;.
\label{lnPwDADB}
\end{equation}
To find the best fit parameters $\bfw$ requires us to minimise the
above probability in the $\bfw$ space.  It is as easy to calculate
this statistic as the standard $\chi^2$, and it can be generalized for
any number of data sets.

Since $\alpha$ and $\beta$ have been eliminated from the analysis by
marginalisation they do not have particular values that can be quoted.
Rather, each value of $\alpha$ and $\beta$ has been considered and
weighted according to the probability of the data given the model.  It
can be shown that the `weights' are $ \alpha_{\rm {eff}} = \frac {
N_{A}} {\chi_{A}^{2}}$ and $\beta_{\rm {eff}} = \frac { N_{B}}
{\chi_{B}^{2}} $, both evaluated at the joint peak.


\begin{thebibliography}{}

\bibitem[]{} Allen S.W., Schmidt R.W., Fabian A.C., 2002, MNRAS, 334, L11
\bibitem[]{} Bahcall, N.A., Ostriker, J.P., Perlmutter, S.,
Steinhardt, P.J., 1999, Science, 284, 148
\bibitem[]{} Bahcall, N.A., Comerford J.M., 2002, ApJ, 565, 5
\bibitem[]{} Bridle, S.L., Eke, V.R., Lahav, O., Lasenby, A.N.,
Hobson, M.P., Cole, S., Frenk, C.S., \& Henry, J.P. 1999, MNRAS, 310,
565
\bibitem[]{} Bridle S.L., 2000, PhD thesis, University of Cambridge
\bibitem[]{} Bridle S.L., Zehavi I., Dekel A., Lahav O., Lasenby A.N.,
Hobson M.P., 2001a, MNRAS, 321, 333
\bibitem[]{} Bridle, S.L., Crittenden R., Melchiorri A., Hobson M.P.,
Kneissl R., Lasenby A.N., 2002, MNRAS, 335, 1193
\bibitem[]{} Burles S., Nollett K.M., Turner M.S., 2001, ApJ, 552,
L1-L6
\bibitem[]{} de Bernardis P. {\it et al.}, 2002, ApJ, 564, 559
\bibitem[]{} Douspis M., Blanchard A., Sadat R., Barlett J.G., Le Dour
M.,2001, A\&A 379, 1
\bibitem[]{} Efron B., 1982, {\it The jackknife, the bootstrap and
other resampling plans} Regional Conference Series in Applied
Mathematics, no. 38, SIAM
\bibitem[]{} Efstathiou G., Bridle S.~L., Lasenby A.~N., Hobson M.~P.,
Ellis R.~S. 1999, MNRAS, 303, L47
\bibitem[]{} Efstathiou G. \& 2dFGRS team, 2002, MNRAS, 330, 29
\bibitem[]{} Eisenstein, D.J., Hu, W., Tegmark, M., 1999, ApJ, 518, 2
\bibitem[]{} Ettori S., Fabian A.C., 1999, MNRAS 305, 834 (EF99)
\bibitem[]{} Ettori S., Allen S.W., Fabian A.C., 2001, MNRAS, 322, 187
\bibitem[]{} Ettori S., 2001, MNRAS, 323, L1
\bibitem[]{} Evrard A.E., Metzler C.A., Navarro J.F., 1996, ApJ, 469,
494
\bibitem[]{} Freedman, W.L., {\it et al.}, 2001, ApJ, 553, 47
\bibitem[]{} Fukugita M., Hogan C.J., Peebles P.J.E., 1998, ApJ 503,
518
\bibitem[]{} Gawiser E., Silk J., 1998, Science, 280, 1405
\bibitem[]{} Hasinger G. {\it et al.}, 2001, A\&A, 365, 45
\bibitem[]{} Henry J.P., 2000, ApJ 545, 565
\bibitem[]{} Hobson M.P., Bridle S.L., Lahav O., 2002, MNRAS, 335, 377
\bibitem[]{} Kitayama T., Suto Y., 1996, ApJ 469, 480
\bibitem[]{} Knox L., Christensen N., Skordis C., 2001, ApJ, 563, 95
\bibitem[]{} Lahav, O., Bridle, S.L., Hobson, M.P., Lasenby, A.L.,
Sodr\'e, L.  2000, MNRAS, 315, 45
\bibitem[]{} Lahav, O., 2001a, in the proceedings of IAU201 {\it New
Cosmological Data and the Values of the Fundamental Parameters}
Manchester 2001, Eds. A. Lasenby and A. Wilkinson, in press,
(astro-ph/0012475)
\bibitem[]{} Lahav, O., 2001b, in the proceedings of {\it
Astrophysical Ages and Time Scales}, Hawaii 2001, ASP conference
series, pg. 617, Eds.  T. von Hippel, C. Simpson and N. Manset
(astro-ph/0105352)
\bibitem{} Lahav O. \& 2dFGRS team, 2002, MNRAS, 333, 961
\bibitem[]{} Lineweaver C. H. 1998, ApJ, 505, L69
\bibitem[]{} Mohr J.J., Mathiesen B., Evrard A.E., 1999, ApJ,517, 627
(MME99)
\bibitem[]{} Netterfield C.B., {\it et al.}, 2002, ApJ, 571, 604
\bibitem[]{} Percival, J.W. \& the 2dFGRS team, 2001,MNRAS, 327, 1297
\bibitem[]{} Press, W.H., 1996, in  {\it Unresolved Problems in
Astrophysics}, Proceedings of Conference in Honor of John Bahcall,
ed. J.P. Ostriker, Princeton: Princeton University Press (astro-ph/9604126)
\bibitem[]{} Pryke C., Halverson N.W., Leitch E.M., Kovac J.,
Carlstrom J.E., Holzapfel W.L., Dragovan M., 2002, ApJ, 568, 46
\bibitem[]{} Shanks T., Allen P.D., Hoyle F., Tanvir N.V., 2002, 
in {\it A New Era in Cosmology}, Eds. Metcalfe N., Shanks T., ASP Conf
Ser, in press, (astro-ph/0208237)
\bibitem[]{} Steigman G., Hata N., Felten J.E. 1999, ApJ, 510, 564S
\bibitem[]{} Turner E.L., Cen R., Ostriker J., 1992, ApJ, 103, 1427 
\bibitem[]{} Wang L., Caldwell R. R., Ostriker J.P., Steinhardt, 1999,
ApJ, 530,17
\bibitem[]{} Webster, M., Bridle, S.L., Hobson, M.P., Lasenby, A.N.,
Lahav, O., \& Rocha, G.  1998, ApJ Lett, 509, L65
\bibitem[]{} Weisskopf M.C., Tanabaum H.D., Van Spebroeck L.P., O'
Dell S.L., 2000, Proc SPIE 4012,2 (astro-ph/0004127)
\bibitem[]{} White S.D.M., Navarro J.F., Evrard A.E., Frenk C.S.,
1993, Nature 366, 429
\end{thebibliography}
\end{document}